\documentstyle[shortnote,twocolumn,epsf]{jpsj2}

\catcode`\@=11

\def\simle{\mathrel{\mathpalette\@versim<}}   
\def\simge{\mathrel{\mathpalette\@versim>}}   
\def\@versim#1#2{\lower2.5pt\vbox{\baselineskip0pt \lineskip-.5pt
   \ialign{$\m@th#1\hfil##\hfil$\crcr#2\crcr\sim\crcr}}}

\catcode`\@=12

\def\runtitle{
}
\def\runauthor{Yukitoshi {\sc Motome} and Nobuo {\sc Furukawa}}

\def\sf{\rm}

\title{
Universality Class of Ferromagnetic Transition in\\
Three-Dimensional Double-Exchange System \\
- O(N) Monte Carlo Study -
}

\author{
Yukitoshi {\sc Motome} and Nobuo {\sc Furukawa}$^{1}$ 
}
\inst{
RIKEN (The Institute of Physical and Chemical Research), 
Wako 351-0198\\
$^{1}$Department of Physics, Aoyama Gakuin University, 
Sagamihara 229-8558
}
\recdate{
}

\abst{
Curie temperature and exponents are studied 
for the three-dimensional double-exchange model. 
Applying the $O(N)$ Monte Carlo algorithm, 
we perform systematic finite-size scaling analyses 
on the data up to $20^3$ sites. 
The obtained values of the critical exponents are 
consistent with those of the Heisenberg universality class, and
clearly distinct from the mean-field values.
}

\kword{
double-exchange system, ferromagnetic transition, universality class,
Monte Carlo, O(N) method, colossal-magnetoresistance manganites
}

\begin{document}
\sloppy
\maketitle


Rediscovery of colossal magnetoresistance (CMR) phenomena
has revived much interests in manganese oxides.
\cite{Ramirez1997,Furukawa1999}
As sample quality and experimental precision are improved,
many experiments have been done 
to examine critical properties
of the ferromagnetic transition in these compounds.
Critical properties, especially critical exponents 
which assign the universality class of the transition, 
are highly important to understand the nature of the transition 
as well as basic physics of the CMR phenomena.
Estimates of critical exponents, however,
are scattered between different experimental results and 
widely range from the Heisenberg-like ones to the mean-field-like ones.
\cite{FurukawaPREPRINT}
There still remains experimental controversy on this issue.

Theoretical estimates of critical exponents 
in the double-exchange (DE) model,
which is considered to be a relevant model in these compounds,
\cite{Zener1951}
would help to reconcile this experimental controversy.
However, it is difficult to clarify critical properties of this transition
because the system is a strongly-correlated electron system;
itinerant electrons interact with localized spins
through a strong Hund's-rule coupling
whose magnitude is much larger than the bandwidth of electrons.
Near the finite-temperature transition,
spin fluctuations are critically enhanced
due to the interplay between itinerant electrons and localized spins.
A sophisticated tool is necessary to handle this many-body problem.

Numerical calculation which includes all the correlation effects
is one of the most promising tools for this problem.
Monte Carlo (MC) method has been applied 
for studying the critical properties of this system.
\cite{Motome2000}
However, most of the MC studies suffered from finite-size effects
since they were limited to small-size systems
due to the rapid increase of the computational cost
as the system size increases.
Systematic analysis on the finite-size effect is indispensable
to examine the critical properties of this transition in detail.

The authors studied this problem by developing a MC algorithm, 
which is called the polynomial expansion Monte Carlo (PEMC) method. 
\cite{Motome1999,Motome2000}
In this method, the computational cost has reduced to $O(N^3)$
from $O(N^4)$ in the conventional MC method 
\cite{Yunoki1998}
($N$ is the system size), 
and the finite-size scaling was applied to results up to $8^3$ sites. 
The problem was also studied by Alonso {\it et al.}
\cite{Alonso2001}
They applied the hybrid MC algorithm in which the computational cost 
is proportional to $O(N^2)$ (they claim it can be reduced to $O(N)$ empirically), 
and estimated the critical exponent from the data up to $12^3$ sites. 
In both studies, the estimates of the exponents are 
consistent with those in the universality class of 
the Heisenberg model with short-range interactions. 
They are, however, not enough to estimate the values precisely and 
to show a clear deviation from the mean-field values. 
Further study supported by an improvement in the algorithm 
is strongly desired for this purpose. 

In this work, we give precise estimates of
critical exponents of the ferromagnetic transition
in the three-dimensional (3D) DE system. 
We apply an improved PEMC algorithm
in which the computational cost scales to $O(N)$,  
\cite{FurukawaPREPRINT2}
and perform the systematic finite-size scaling for 
the data from $6^3$ to $20^3$ sites.


The DE model considered here is defined by the Hamiltonian
\cite{Zener1951}
\begin{equation}
{\cal H} = -t \sum_{\langle ij \rangle, \sigma}
( c_{i \sigma}^\dagger c_{j \sigma} + {\rm h.c.} )
- J_{\rm H} \sum_i {\mib \sigma}_i \cdot {\mib S}_i,
\label{eq:H}
\end{equation}
where the first term describes the nearest-neighbor hopping
of itinerant electrons on 3D cubic lattices and 
the second term is for the Hund's-rule coupling 
between the electrons and the localized spins ${\mib S}$.
For simplicity, we take the limits of 
$J_{\rm H} \rightarrow \infty$ and
$S = |{\mib S}| \rightarrow \infty$ (classical spins)
in the following calculations.


We study the ferromagnetic transition in model (\ref{eq:H})
by using an improved PEMC method.
In this method, we introduce effective truncations
for the vector-matrix product and the trace operation,
which reduce the computational cost for one MC update to $O(N)$.
Readers are referred to Ref. \citen{FurukawaPREPRINT2}
for the details of this algorithm.
This technique enables us to study up to $20^3$ sites
within a reasonable time on parallel computers.
We choose periodic boundary conditions and
have typically run $4000$ MC samplings for measurements
after thermalization with $1000$ MC steps.
The doping concentration is fixed at quarter filling,
namely, $0.5$ electron per site on average.
The energy unit is the half-bandwidth of noninteracting case, 
$W = 6t = 1$.

\begin{figure}
\epsfxsize=6.5cm
\centerline{\epsfbox{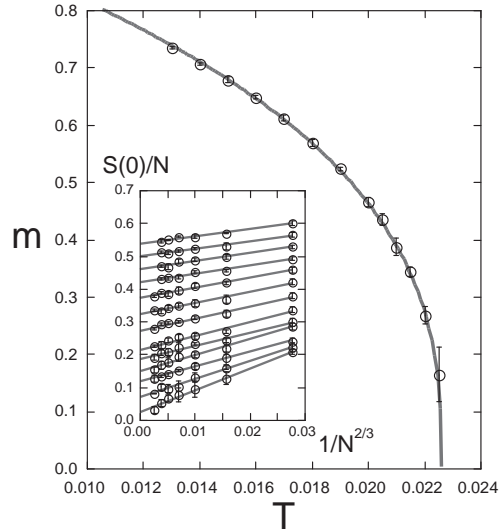}}
\caption{
Temperature dependence of the magnetization
on the thermodynamic limit.
The gray curve is the least-squares-fit 
to $m \propto (T-T_{\rm c})^\beta$.
Inset: System-size extrapolation of the spin structure factor.
The data are for $T=0.013,\cdot\cdot\cdot,0.0225$
from top to bottom.
}
\label{fig:MvsT}
\end{figure}

Figure \ref{fig:MvsT} shows the temperature dependence
of the magnetization $m = [S({\mib k}=0)/N]^{1/2}$.
Here $S({\mib k})$ is the spin structure factor
which is defined as 
$S({\mib k}) = \sum_{ij} \langle {\mib S}_i \cdot
{\mib S}_j \rangle \exp{({\rm i} {\mib k} \cdot {\mib r}_{ij})} / N$,
where the bracket denotes the thermal average 
for the grand canonical ensemble.
The magnetization data are the thermodynamic-limit values
which are estimated by the system-size extrapolation of
$S(0)/N$ as shown in the inset.
We fit the magnetization data by assuming
the scaling relation $m \propto (T_{\rm C} - T)^\beta$
(the gray curve in the figure)
to estimate the value of $T_{\rm C}$ and 
the critical exponent $\beta$.
The fit gives $T_{\rm C} = 0.0226(2)$ and $\beta = 0.36(1)$.
(Errors in the last digit are shown in parentheses.)
The estimate of $\beta$ is consistent with
the Heisenberg one $\beta = 0.365$ and
definitely distinct from the mean-field one $\beta = 0.5$.

\begin{figure}
\epsfxsize=8cm
\centerline{\epsfbox{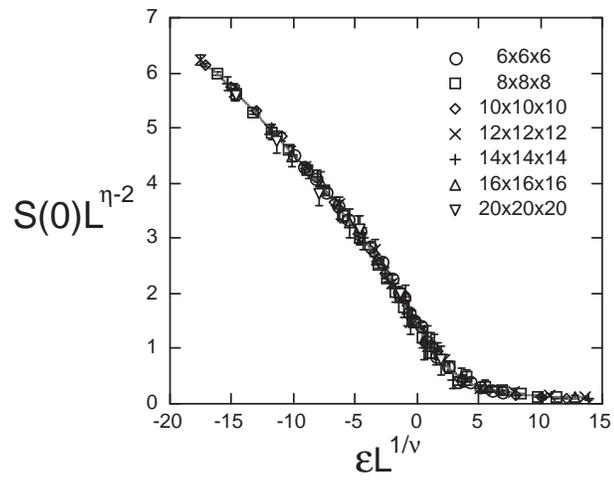}}
\caption{
The best-fit result of the finite-size scaling plot for 
the Monte Carlo data.}
\label{fig:scaling}
\end{figure}

We also examine the finite-size scaling plot
for $S(0)$ to determine the critical exponents.
The scaling form is given by
$S(0) L^{\eta-2} = f( \varepsilon L^{1/\nu} )$
which is derived based on the hyperscaling hypothesis.
Here $\varepsilon = (T-T_{\rm C})/T_{\rm C}$, $L = N^{1/3}$
and $f$ is the scaling function.
We plot $S(0) L^{\eta-2}$ as a function of $\varepsilon L^{1/\nu}$,
and optimize the values of ($T_{\rm C}, \eta, \nu$)
so that all the MC data for different $L$ and $T$
converge to a single curve.
\cite{Motome2001}
Figure \ref{fig:scaling} shows the optimized fit.
This fit gives the estimates as
$T_{\rm C} = 0.0222(+5,-3)$, 
$\nu = 0.60(+13,-10)$ and $\eta = 0.07(+34,-7)$.
By using the relation $\beta = \nu (1+\eta) / 2$ with
the obtained distribution of $\nu$ and $\eta$,
we obtain the estimate $\beta = 0.32(+10,-6)$.
Both estimates of $T_{\rm C}$ and $\beta$ are consistent with
the estimates in Fig.~\ref{fig:MvsT}.
We confirmed that the mean-field exponents $\nu=1/2$ and $\eta=0$
do not give a satisfactory convergence in the scaling fit.


These results indicate that
the ferromagnetic transition in the DE model (\ref{eq:H})
belongs to the universality class of the Heisenberg spin system
with short-range interactions rather than to that of the mean-field solution.
This suggests that the critical exponents in the Heisenberg universality class 
should be observed in manganites
when the DE mechanism plays a key role in the transition
as widely believed.
Other additional elements which are not included
in the simple model (\ref{eq:H}), 
such as the electron-lattice coupling,
may not affect the universality class
because the fluctuations through the additional degrees of freedom 
will make correlations more short-ranged.


To summarize, we have investigated the critical exponents
in the ferromagnetic transition 
in the three-dimensional double-exchange model
by using the improved Monte Carlo method.
The finite-size scaling analysis up to $20^3$ sites
indicates that the transition belongs to
the universality class of the Heisenberg spin system
with short-range interaction and is distinct from the mean-field one.
In manganese oxides, we expect
this Heisenberg universality class to be observed
if the double-exchange mechanism plays a major role in the transition.
Our results give a theoretical background
to reconcile the experimental controversy
on the estimates of the critical exponents.


The authors thank H. Nakata for helpful support
in developing parallel-processing systems.
The computations have been performed mainly 
using the facilities in the AOYAMA+ project
(http://www.phys.aoyama.ac.jp/ \\
\~{}aoyama+).
This work is supported by  ``a Grant-in-Aid from
the Ministry of Education, Culture, Sports, Science and Technology''.




\end{document}